\documentclass{article}
\usepackage{amsmath,amsthm,amssymb}
\usepackage{url}

\begin{document}
\title{Bootstrapping the Long Tail in Peer to Peer Systems}
\author{Bernardo A. Huberman and Fang Wu\\HP Labs, Palo Alto, CA 94304}
\maketitle

\begin{abstract}
We describe an efficient incentive mechanism for P2P systems that
generates a wide diversity of content offerings while responding
adaptively to customer demand. Files are served and paid for through
a parimutuel market similar to that commonly used for betting in
horse races. An analysis of the performance of such a system shows
that there exists an equilibrium with a long tail in the
distribution of content offerings, which guarantees the real time
provision of any content regardless of its popularity.
\end{abstract}

\pagebreak
\section{Introduction}
\medskip

The provision of digitized content on-demand to millions of users
presents a formidable challenge. With an ever increasing number of
fixed and mobile devices with video capabilities, and a growing
consumer base with different preferences, there is a need for a
scalable and adaptive way of delivering a diverse set of files in
real time to a worldwide consumer base.

Providing such varied content presents two problems. First, files
should be accessible in such a way that the constraints posed by
bandwidth and the diversity of demand is met without having to
resort to client server architectures and specialized network
protocols. Second, as new content is created, the system ought to be
able to swiftly respond to new demand on specific content,
regardless of its popularity. This is a hard constraint on any
distributed system, since providers with a finite amount of memory
and bandwidth will tend to offer the most popular content, as is the
case today with many peer-to-peer systems.

The first problem is naturally solved by peer to peer networks,
where each peer can be both a consumer and provider of the service.
Peer to peer networks, unlike client server architectures,
automatically scale in size as demand fluctuates, as well as being
able to adapt to system failures. Examples of such systems are
Bittorrent \cite{cohen-03} and Kazaa, who account for a sizable
percentage of all the use of the Internet. Furthermore, new services
like the BBC IMP, (\url{http://www.bbc.co.uk/imp/}) show that it is
possible to make media content available through a peer-to-peer
system while respecting digital rights.

It is the second problem, that of an adaptable and efficient system
capable of delivering any file, regardless of its popularity, that
we now solve. We do so by creating an implementable incentive
mechanism that ensures the existence of a diverse set of offerings
which is in equilibrium with the available supply and demand,
regardless of content and size. Moreover, the mechanism is such that
it automatically generates the long tail of offerings which has been
shown to be responsible for the success of a number of online
businesses such as Amazon or eBay \cite{anderson-05}. In other
words, while the system delivers favorite mainstream content, it can
also provide files that constitute small niche markets which only in
the aggregate can generate large revenues.

In what follows we describe an efficient incentive mechanism for
P2P systems that generates a wide diversity of content offerings
while responding adaptively to customer demand. Files are served
and paid for through a parimutuel market similar to that commonly
used for betting in horse races. An analysis of the performance of
such a system shows that there exists an equilibrium with a long
tail in the distribution of content offerings, which guarantees
the real time provision of any content regardless of its
popularity. In our case, the bandwidth fraction of a given file
offered by a server plays the role of the odds, the bandwidth
consumed corresponds to bettors, the files to horses, and the
requests are analogous to races.

An interesting consequence of this mechanism is that it solves in
complete fashion the free riding problem that originally plagued
P2P systems like Gnutella \cite{adar-huberman} and that in milder
forms still appears in other such systems. The reason being that
it transforms the provision of content from a public good into a
private one.

We then analyze the performance of such a system by making a set of
assumptions that are first restrictive and are then relaxed so as to
make them correspond to a realistic crowd of users. We show that in
all these cases there exists an equilibrium in which the demand for
any file can be fulfilled by the system. Moreover this equilibrium
exhibits a robust empirical anomaly which is responsible for
generating a very long tail in the distribution of content
offerings. We finally discuss the scenario where most of the servers
are bounded rational and show that it is still possible to achieve
an optimum equilibrium. We conclude by summarizing our results and
discussing the feasibility of its implementation.

\section{The system and its incentive mechanism}
\medskip

Consider a network-based file exchange system consisting of three
types of traders: content provider, server, and downloader or
user. A content provider supplies---at a fixed price per file---a
repertoire of files to a number of people acting as peers or
servers. Servers then selectively serve a subset of those files to
downloaders for a given price. In a peer-to-peer system a
downloader can also, and often does, act as a server.

If the files are typically large in size, a server can only afford
to store and serve a relatively small subset of files. It then faces
the natural problem of choosing an optimal (from the point of view
of maximizing his utility) subset of files to store so as to sell
them to downloaders.

Suppose that the system charges each downloader a flat fee for
downloading any one file (as in~Apple's iTunes music store), which
we normalize to one. Since many servers can help distribute a single
file, this unit of income has to be allocated to the servers in ways
that will incentivize them to always respond to a changing demand.

In order to do so, consider the case where there are $m$ servers and
$n$ files. Let $b_{ij}$ be the effective bandwidth of server $i$
serving file $j$, normalized to \begin{equation} \sum_{i,j}
b_{ij}=1.
\end{equation}
Also, denote the bandwidth fraction of file $j$ by \begin{equation}
\pi_j=\sum_k b_{kj}.
\end{equation}

Suppose that when a downloader connects to the system, it starts
downloading different parts of the file simultaneously from all
available servers that have it. When it finishes downloading, it
will have received a fraction of the file $j$ \begin{equation}
\label{eq:prop-share} q_{ij}=\frac{b_{ij}}{\sum_k b_{kj}} =
\frac{b_{ij}} {\pi_j} \end{equation} from server $i$. Our mechanism
prescribes that \emph{the system should should pay an amount
$q_{ij}$ to server $i$ as its reward for serving file $j$}.

Now consider the case when server $i$'s reserves an amount of
bandwidth $b_{ij}$ as his ``bid'' on file $j$. Because we have
normalized the total bandwidth and the total reward for serving one
request both to one, the proportional share allocation scheme
described by Eq.~(\ref{eq:prop-share}) can be interpreted as
redistributing the total bid to the ``winners'', in proportion to
their bids. Thus our payoff structure is similar to that of a
pari-mutuel horse race betting market, where the $\pi_j$ can be
regarded as the odds, the bandwidth corresponds to bettors, the
files to horses, and the requests are analogous to races. It is
worth pointing out however, that in a real horse race all players
who have placed a bet on the winning horse receive a share of the
total prize, whereas in our system only those players that kept the
"winning" file and also had a chance to serve it get paid. In spite
of this difference it is easy to show that when rewritten in terms
of expected payoffs, the two mechanisms behave in similar fashion.
\section{The solution}

\subsection{Rational servers with static strategies and known\\ download rates}

In this section we make three simplifying assumptions. While not
realistic they serve to set the framework that we will utilize later
on to deal with more realistic scenarios. First, every server is
rational in the sense that he chooses the optimal bandwidth
allocation that maximizes his utility, whose explicit form will be
given below. Second, every server's allocation strategy is static,
i.e.~the $b_{ij}$'s are independent of time. Third, we assume that
each file $j$ is requested randomly at a rate $\lambda_j>0$ that
does not change with time, and these rates are known to every
server.

Consider a server $i$ with the following standard additive form of
utility:
\begin{equation}
U=\mathbb E \left[ \int_0^\infty e^{-\delta t} u(t) dt \right],
\end{equation} where $u(t)$ is his income density at time $t$, and
$\delta>0$ is his future discount factor. Let $X_{j1}$ be the
(random) time that file $j$ is requested for the first time, let
$X_{j2}$ be the time elapsed between the first request and the
second request, and so on. According to our parimutuel reward
scheme, server $i$ receives a lump-sum reward $b_{ij}/\pi_j$ from
every such request, at times $X_{j1}, X_{j1}+X_{j2}$, etc. Thus, the
server $i$'s total utility is given by \begin{equation}
\label{eq:U-1} U=\sum_j \frac{b_{ij}}{\pi_j} \sum_{l=1}^\infty
\mathbb E [e^{-\delta \sum_{k=1}^l X_{jk}}] \equiv \sum_j
\frac{b_{ij}}{\pi_j} u_j.
\end{equation}
The sum of expectations in Eq.~(\ref{eq:U-1}) (denoted by $u_j$) can
be calculated explicitly. Because the $X_{jk}$'s are i.i.d.~random
variables with density $\lambda_j^{-1} \exp(\lambda_j x)$, we have
\begin{equation} u_j = \mathbb E[e^{-\delta X_{j1}}] \left(1+
\sum_{l=2}^\infty \mathbb E [e^{-\delta \sum_{k=2}^l X_{jk}}]
\right) = \frac {\lambda_j}{\lambda_j+\delta} (1+u_j).
\end{equation}
Solving for $u_j$, we then find
\begin{equation}
u_j=\frac {\lambda_j}\delta.
\end{equation}
If we let $\lambda = \sum_j \lambda_j$ be the total request rate and
$p_j=\lambda_j/\lambda$ be the probability that the next request
asks for file $j$, then we can also write \begin{equation} u_j=\frac
\lambda\delta p_j.
\end{equation}
Plugging this back into Eq.~(\ref{eq:U-1}), we obtain
\begin{equation} U= \frac \lambda \delta \sum_j \frac{p_j
b_{ij}}{\pi_j}.
\end{equation}

Since we assume that server $i$ is rational, he will allocate
$b_{ij}$ in a way that it solves the following optimization problem:
\begin{equation}
\label{eq:opt} \max_{(b_{ij})_{j=1}^n\in \mathbb R^n_+} \sum_j
\frac{p_j b_{ij}}{\sum_k b_{kj}} \quad \text{subject to } \sum_j
b_{ij} \le b_i.
\end{equation}
Thus we see that the servers are playing a \emph{finite budget
resource allocation game}. This type of game has been studied
intensively, and a Nash equilibrium has been shown to exist under
mild assumptions \cite{shapley-shubik-77, zhang-05}. In such an
equilibrium, the players' utility functions are strongly competitive
and in spite of a possibly large utility gap, the players behave in
almost envy-free fashion, i.e.~each player believes that that no
other player has received more than they have.

\subsection{Rational servers with static strategies and unknown request rates} We now relax some of the assumptions made above so as to deal with a more realistic case.

It is usually hard to find out the accurate request rate for a given
file, especially at the early stages when there is no historical
data available. Thus it makes more sense to assume that every server
$i$ holds a \emph{subjective belief} about those request rates. Let
$p_{ij}$ be server $i$'s subjective probability that the next
request is for file $j$. Then server $i$ believes that file $j$ will
be requested at a rate $\lambda_{ij} = \lambda p_{ij}$.
Eq.~(\ref{eq:opt}) then becomes \begin{equation}
\max_{(b_{ij})_{j=1}^n\in \mathbb R^n_+} \sum_j \frac{p_{ij}
b_{ij}}{\sum_k b_{kj}} \quad \text{subject to } \sum_j b_{ij} \le
b_i.
\end{equation}
which is still a finite budget resource allocation game as
considered in the previous section.

It is interesting to note that when $m$ is large, $b_{ij}$ is small
compared to $\pi_j = \sum_k b_{kj}$, so that $\pi_j$ can be treated
as a constant. In this case, the optimization problem can be well
approximated by \begin{equation} \label{eq:consensus}
\max_{(b_{ij})_{j=1}^n\in \mathbb R^n_+} \sum_j \frac{p_{ij}
b_{ij}}{\pi_j} \quad \text{subject to } \sum_j b_{ij} \le b_i.
\end{equation}
Thus, user $i$ should use all his bandwidth to serve those files $j$
with the largest ratio $p_{ij}/\pi_j$.

This scenario (\ref{eq:consensus}) corresponds to the so-called
\emph{parimutuel consensus} problem, which has been studied in
detail. In this problem a certain probability space is observed by a
number of individuals, each of which endows it with their own
subjective probability distributions. The issue then is how to
aggregate those subjective probabilities in such a way that they
represent a good consensus of the individual ones. The parimutuel
consensus scheme is similar to that of betting on horses at a race,
the final odds on a given horse being proportional to the amount bet
on the horse. As shown by Eisenberg and Gale
\cite{eisenberg-gale-58}, an equilibrium then exists such that the
bettors as a group maximize the weighted sum of logarithms of
subjective expectations, with the weights being the total bet on
each horse.

Moreover a number of empirical studies of parimutuel markets
\cite{thaler-ziemba-88} have shown that they do indeed exhibit a
high correlation between the subjective probabilities of the bettors
and the objective probabilities generated by the racetracks. Equally
interesting for our purposes is the existence of a robust empirical
anomaly called \emph{the favorite-longshot bias}
\cite{thaler-ziemba-88}. The anomaly shows that favorites win more
frequently than the subjectives probabilities imply, and longshots
less often. Besides implying that favorites are better bets than
long shots, this anomaly ensures the existence of the long tail,
populated by those files which while not singly popular, in
aggregate are responsible for a large amount of the traffic in the
system.

\subsection{Rational servers with a dynamic strategy}

We now consider the case where the rate at which files are requested
can change with time. Because of this, each server has to actively
adjust its bandwidth allocation to adapt to such changes. As we have
seen in the last section, user $i$ has an incentive to serve those
files with large values of $p_{ij}/\pi_j$. Recall that $\pi_j(t)$ is
just the fraction of total bandwidth spent to serve file $j$ at time
$t$, which in principle can be estimated from the system's
statistics. Thus it would be useful to have the system frequently
broadcast the real-time $\pi_j$ to all servers so as to help them
decide on how to adjust their own allocations of bandwidth.

From Eq.~(\ref{eq:prop-share}) we see that, by serving file $j$,
user $i$'s expected per bandwidth earning from the next request is
\begin{equation} \frac{p_j q_{ij}} {b_{ij}} = \frac{p_j}{\pi_j}.
\end{equation}
Hence a user will benefit most by serving those files with the
largest ``$p/\pi$ ratio''. However, as soon as a given user starts
serving file $j$, the corresponding $p/\pi$ ratio decreases. As a
consequence, the system self-adapts to the limit of uniform $p/\pi$
ratios. If the system is perfectly efficient, we would expect that
\begin{equation} \frac {p_j}{\pi_j} = \text{constant}.
\end{equation}
Because $p_j$ and $\pi_j$ both sum up to one, this implies that
\begin{equation} \pi_j=p_j, \end{equation} or \begin{equation}
\sum_k b_{kj} = \frac {\lambda_j}\lambda \propto \lambda_j.
\end{equation}
In other words, the total bandwidth used to serve a file is
proportional to the file's request rate.

This result has interesting implications when considering the social
utility of the downloaders. Recently, Tewari and Kleinrock
\cite{tewari-kleinrock-05} have shown that in a homogeneous network
the average download time is minimized when $\sum_k b_{kj} \propto
\lambda_j$. This implies that in the perfectly efficient limit, our
mechanism maximizes the downloaders' social utility, which is
measured by their average download times.

Since in reality a market is never perfectly efficient, the above
analysis only makes sense if the characteristic time it takes for
the system to relax back to uniformity from any disturbance is
short. As a concrete example, consider a new file $j$ released at
time 0, being shared by only one server. Suppose that every
downloader starts sharing her piece of the file immediately after
downloading it. Because there are few servers serving the file but
many downloaders requesting the file, for very short times
afterwards the upload bandwidth will be fully utilized. That is,
during time $dt$, an amount $\pi_j(t) dt$ of data is downloaded
and added to the total upload bandwidth immediately. Hence we have
\begin{equation}
d\pi_j(t) = \pi_j(t) dt.
\end{equation}
which implies that $\pi_j(t)$ grows exponentially until
$\pi_j(T)\sim p_j$. Solving for $T$, we find \begin{equation} T\sim
\log \left( \frac {p_j}{\pi_j(0)} \right).
\end{equation}
Thus the system reaches uniformity in logarithmic time, a signature
of its high efficiency.

\subsection{Servers with bounded rationality}

So far we have assumed that all servers are rational, so that they
will actively seek those files that are most under-supplied so as
to serve them to downloaders. In reality however, while some
servers do behave rationally, a lot others do not.
 This is because even a perfectly rational server sometimes can make
  wrong decisions as to which files to store because his subjective
   probability estimate of what is in demand can be inaccurate. Also, such
 a bounded-rational server can at times be too lazy to adjust his bandwidth allocation, so that he will keep serving whatever he has, and at other times he might simply imitate other servers'
behavior by choosing to serve the popular files. In all these cases
we need to consider whether or not the lack of full rationality will
lead to equilibrium on the part of the system.

As a simple example, assume there are only two files, A and B. Let
$p=\lambda_A/\lambda$ be file A's real request probability, and let
$1-p$ be file B's real request probability. Suppose the servers are
divided into two classes, with $\alpha$ fraction rational and
$1-\alpha$ fraction irrational, arriving one by one in a random
order. Each rational server's subjective probability in general can
be described by an identically distributed random variable $P_t \in
[0,1]$ with mean $p$.  Then with probability $\mathbb P[P_t >
\pi(t)]$ he will serve file A, and with probability $\mathbb
P[P_t<\pi(t)]$ he will serve file B. In order to carry out some
explicit calculation below, we consider the simplest choice of
$P_t$, namely a Bernoulli variable \begin{equation} \mathbb
P[P_t=1]=p, \quad \mathbb P[P_t=0]=1-p.
\end{equation}
(Clearly $\mathbb E[P_t]=p$, so the subjective probabilities are
accurate on average.) It is easy to check that under this choice a
rational server chooses A with probability $p$ and B with
probability $1-p$.

On the other hand, consider the situation where an irrational server
chooses an existing server at random and copies that server's
bandwidth allocation. That is, with probability $\pi(t)$ an
irrational server will choose file A.\footnote{This assumption can
also be interpreted as follows. Suppose a downloader starts serving
his files immediately after downloading, but never initiates to
serve a file. (This is the way a non-seed peer behaves within
Bittorrent.) Then the probability that he will serve file $j$ is
exactly the probability that he just downloaded file $j$, which is
$\pi_j(t)$.}

From these two assumptions we see that
\begin{equation}
\mathbb P[\text{server $t$ serves A}]= \alpha p + (1-\alpha) \pi(t),
\end{equation} and \begin{equation} \mathbb P[\text{server $t$
serves B}]= \alpha (1-p) + (1-\alpha) (1-\pi(t)).
\end{equation}

The stochastic process described by the above two equations has been
recently  studied in the context of choices among technologies for
which evidence of their value is equivocal, inconclusive, or even
nonexistent \cite{bendor-huberman-wu-05}. As was shown there, the
dynamics generated by such equations leads to outcomes that appear
to be deterministic in spite of being governed by a stochastic
process. In the context of our problem this means that when the
objective evidence for the choice of a particular file is very weak,
any sample path of this process quickly settles down to a fraction
of files downloaded that is not predetermined by the initial
conditions: ex ante, every outcome is just as (un)likely as every
other. Thus one cannot ensure an equilibrium that is both optimum
and repeatable.

In the opposite case, when the objective evidence is strong, the
process settles down to a value that is determined by the quality of
the evidence. In both cases the proportion of files downloaded never
settles into either zero or one.

In the general case that we have been considered, there are always a
number of servers that will behave in bounded rational fashion and a
few that are perfectly rational. Specifically, when $\alpha>0$,
which corresponds to the case where a small number of servers are
rational, the $\pi(t)$ will converge to $p$ in the long time limit.
That is, a small fraction of rational servers is enough for the
system to reach an optimum equilibrium. However, it is worth
pointing out that since the characteristic convergence time diverges
exponentially in $1/\alpha$, the smaller the value of alpha
$\alpha$, the longer it will take for the system to reach such an
optimum state.

\section{Conclusion}
\medskip

In this paper we a peer-to-peer system with an incentive mechanism
that generates diversity of offerings, efficiency and adaptability
to customer demand. This was accomplished by having a pricing
structure for serving files that has the structure of a parimutuel
market, similar to those commonly used in horse races, where the the
bandwidth fraction of a given file offered by a server plays the
role of the odds, the bandwidth corresponds to bettors, the files to
horses, and the requests are analogous to races. Notice that this
mechanism completely solves the free riding problem that originally
plagued P2P systems like Gnutella and that in milder forms still
appears in other such systems.

We then analyze the performance of such a system by making a set of
assumptions that are first restrictive but are then relaxed so as to
make the system respond to a realistic crowd. We show that in all
these cases there exists an equilibrium in which the demand for any
file can be fulfilled by the system. Moreover this equilibrium is
known to exhibit a robust empirical anomaly, that of the
\emph{favorite-longshot bias}, which in our case will generate a
very  long tail in the distribution of offerings. We finally
discussed the scenario where most of the servers are bounded
rational and showed that it is still possible to achieve an optimum
equilibrium if a few servers can act rationally.

The implementation of mechanism is completely feasible with present
technologies. The implementation of a prototype will also help study
the behavior of both providers and users within the context of this
parimutuel market. Given its feasibility, and with the addition of
DRM and a payment system, it offers an interesting opportunity for
the provision of legal content with a simple pricing structure that
ensures that unusual content will always be available along with the
more traditional fare.

\paragraph{Acknowledgements.}

We benefited from discussions with Eytan Adar, Tad Hogg and Li
Zhang.

\pagebreak

\end{document}